\documentclass[a4paper,conference]{IEEEconf_ICPR2022/IEEEtran}
\IEEEoverridecommandlockouts
\ifCLASSINFOpdf
\else
\fi
\usepackage{graphicx} 
\usepackage{multirow} 
\usepackage{multicol} 
\usepackage{booktabs} 
\usepackage{amsmath}
\usepackage{amsfonts}
\usepackage{arydshln} 
\usepackage{float}
\usepackage{color, colortbl}
\usepackage[pagebackref,breaklinks,colorlinks]{hyperref}
\usepackage{etoolbox}
\makeatletter
\patchcmd{\@makecaption}
  {\scshape}
  {}
  {}
  {}
\makeatletter
\patchcmd{\@makecaption}
  {\\}
  {.\ }
  {}
  {}
\makeatother
\def\tablename{Table}
\def\figurename{Figure}

\definecolor{red}{rgb}{0.99, 0.25, 0.33}
\definecolor{brickred}{rgb}{0.8, 0.25, 0.33}
\definecolor{iceberg}{rgb}{0.44, 0.65, 0.82}
\definecolor{darkturquoise}{rgb}{0.0, 0.81, 0.82}
\definecolor{dodgerblue}{rgb}{0.12, 0.56, 1.0}
\definecolor{darkcyan}{rgb}{0.0, 0.55, 0.55}

\hypersetup{colorlinks=true}
\hypersetup{linktoc=all}
\hypersetup{citecolor=darkcyan}
\hypersetup{linkcolor=brickred}
\hypersetup{urlcolor=darkcyan}

\usepackage[capitalize]{cleveref}
\crefname{section}{Section}{Secs.}
\Crefname{section}{Section}{Sections}
\Crefname{table}{Table}{Tables}
\crefname{table}{Table}{Tabs.}
\crefname{figure}{Figure}{Figs.}
\Crefname{figure}{Figure}{Figures}

\newcommand{\parahead}[1]{\noindent\textbf{#1.} \enskip}

\newcommand{\edit}[1]{\textcolor{black}{#1}}

\newcommand{\editr}[1]{\textcolor{black}{#1}}

\begin{document}

\title{Automatic Detection of Noisy Electrocardiogram Signals without Explicit Noise Labels}




\author{\IEEEauthorblockN{Radhika Dua\IEEEauthorrefmark{2}\qquad
Jiyoung Lee\IEEEauthorrefmark{2}\qquad
Joon-myoung Kwon\IEEEauthorrefmark{3}\qquad
Edward Choi\IEEEauthorrefmark{2}} \vspace{2mm}
\IEEEauthorblockA{\IEEEauthorrefmark{2}KAIST, Republic of Korea\\
\IEEEauthorblockA{\IEEEauthorrefmark{3}Medical AI Inc., Republic of Korea\\
\tt\small\{radhikadua, jiyounglee0523, edwardchoi\}@kaist.ac.kr}
\tt\small cto@medicalai.com}}


%


\maketitle

\begin{abstract}
Electrocardiogram (ECG) signals are beneficial in diagnosing cardiovascular diseases, which are one of the leading causes of death. However, they are often contaminated by noise artifacts and affect the automatic and manual diagnosis process. Automatic deep learning-based examination of ECG signals can lead to inaccurate diagnosis, and manual analysis involves rejection of noisy ECG samples by clinicians, which might cost extra time. 
To address this limitation, we present a two-stage deep learning-based framework to automatically detect the noisy ECG samples. Through extensive experiments and analysis on two different datasets, we observe that the deep learning-based framework can detect slightly and highly noisy ECG samples effectively. We also study the transfer of the model learned on one dataset to another dataset and observe that the framework effectively detects noisy ECG samples.
\end{abstract}

\IEEEpeerreviewmaketitle

\section{Introduction}
\label{sec:introduction}
Electrocardiogram (ECG) signals provide useful information to clinicians in examining a patient's health status. 
However, ECG signals are often corrupted by various kinds of noise artifacts, including baseline wander, power line interference, muscle artifact, and instrument noise \cite{satija2018new, satija2018new2}. 
They are generally categorized into Level $1$ (clean samples), Level $2$ (slightly noisy samples), and Level $3$ (very noisy samples). 
The presence of these noises in ECG signals are obstacles in diagnosing a patient's status and sometimes makes it impossible to distinguish the basic information such as from which lead the given ECG signal was measured.
Moreover, these noise artifacts degenerate the performance of deep learning models employed on ECG signals because the noise impedes the models in learning key features from the given data \cite{sukor}.
Therefore, it is essential to detect and remove noisy ECG signals before they flow into deep learning models or to clinicians \cite{udit}.

Despite the significance of the automatic detection of noisy ECG signals, not much research has been conducted to tackle this issue. 
Previous research mainly focused on classifying the type of noise that exists in the given signal \cite{udit2, udit3}. These work generate a synthetic dataset in which specific noise types are injected into clean ECG signals which is unnatural and rare in the real-world \cite{PE}.  
In the very early stage, statistical analysis from the data, such as extracted features from QRS peaks, pulse portions, and RR intervals, has been used to detect noisy signals \cite{Orphanidou}. Some work relies on decomposition techniques, like Independent Component Analysis (ICA) and
Empirical Mode Decomposition (EMD) \cite{Yoon, Kuzilek, Jinseok}. However, there has not been much research done on detecting noisy ECG signals using deep learning models \cite{Rodrigues}.  

To this end, we present a two-stage deep learning-based framework to automatically detect the noisy ECG signals, which includes the detection of Level $2$ and Level $3$ ECG samples.
First, we train a Convolutional Auto-encoder (CAE) model only on clean ECG signals that learns to reconstruct the signal.
Then we obtain the feature representation of ECG samples from the latent space of the CAE and use a cluster-conditioned method in the feature space based on Mahalanobis distance \cite{mahala18} to detect the noisy ECG samples.

We conduct a suite of experiments on two datasets that show the promise of the presented approach for the detection of noisy ECG signals. 
We also conducted extensive experiments to study whether the presented approach, trained on
a dataset, can be used to detect noisy samples
in other datasets for which we have limited clean (Level $1$)
samples. We observe that the learned model well transfers to another challenging dataset by finetuning on a small subset of the dataset. To the best of our knowledge, this is the first such effort to use a deep learning-based method to automatically detect noisy ECG samples.

\begin{figure*}[t]
\centering
     \includegraphics[width=0.95\textwidth]{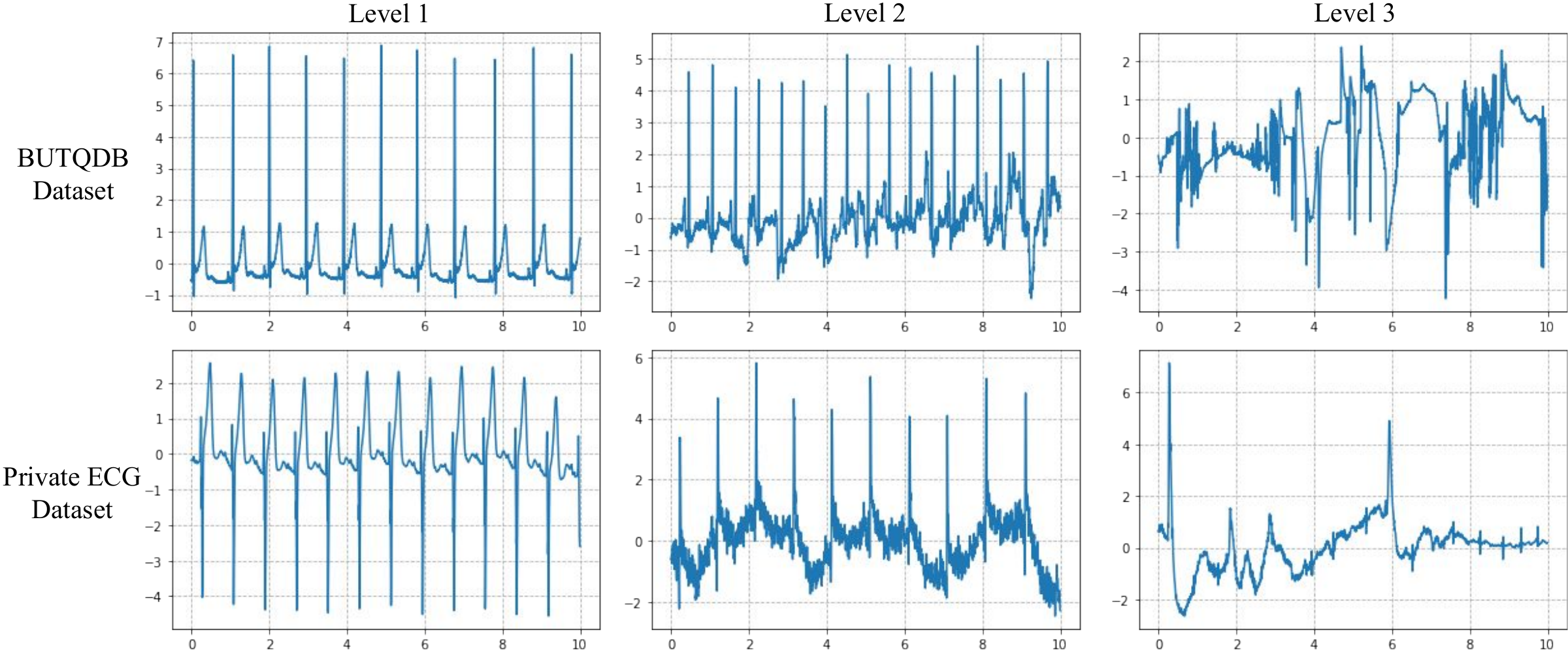}
     \vspace{-4mm}
     \caption{Examples of clean (Level $1$) and noisy (Level $2$ and Level $3$) samples from BUTQDB and Private ECG datasets. Level $3$ samples look much different from Level $1$ samples compared to the Level $2$ samples.}
     \label{fig:ecg_samples}
\end{figure*}

\section{Related Work}
\label{related_work}
In the very early stage of detecting noisy signals, statistical analysis based on the given data, such as extracted features from QRS peaks, RR intervals, has been widely used \cite{Orphanidou}.
Also, some works implement frequency and time domain filters, \textit{e.g.}, wavelet thresholding technique, to detect noisy ECG signals \cite{Donoho, Mohammad, Sayadi}.

Recently, decomposition techniques, like Independent Component Analysis (ICA) and
Empirical Mode Decomposition (EMD), emerge as a new method to detect noisy ECG signals \cite{Yoon, Kuzilek, Jinseok}. 
Adaptive filters, which apply Least Mean Square \cite{Thakor, Guohua}, Bayesian Filters \cite{Sameni, Panigrahy}, are also employed to detect noisy signals.
These adaptive filters have shortcoming in that they need prior training of the data to acquire the best parameters to reconstruct ECG from noisy data. 

Although deep learning
is gaining popularity,
there is limited research related to denoising or removing noisy ECG signals \cite{Rodrigues}. In this paper, we implement a deep learning model to automatically detect noisy signals. 

\section{Noisy ECG signals detection}
\label{task}

 ECG signals are often corrupted by different types of noise artifacts. We refer to such samples as Noisy samples and the task of detecting such samples as Noisy ECG signals detection. These noisy samples are divided into two broad categories, Level $2$ and Level $3$, based on the amount of noise present in the ECG sample (refer to \cref{sec:results} for more details on datasets). Noisy ECG signals detection task solves two purposes: (i) it enhances the reliability of real-world decision making systems by detecting such noisy samples instead of producing incorrect diagnoses; (ii) it reduces clinicians' efforts as they can focus on clean samples for examining a patient's health status.

We now formally define the task.
We consider a dataset composed of clean ECG samples drawn i.i.d. from some underlying data distribution $P_\mathbf{X}$ and train an auto-encoder $g(\mathbf{x})$ on $P_\mathbf{X}$. 
Let $f : \mathbf{x} \rightarrow \mathcal{Z}$, where $\mathcal{Z} \in \mathbb{R}^{m}$, denote the feature extractor parameterized by the encoder of the auto-encoder model $g$, which maps a sample, $\mathbf{x}$, from the input space to the $m$-dimensional latent space $\mathcal{Z}$. 
Our objective is to extract the features of the input sample $\mathbf{x}$ using $f$ and design an approach to detect whether the input sample is noisy or clean.

\begin{table*}[t]
\footnotesize
\centering
\resizebox{\linewidth}{!}{
\begin{tabular}{l  c c  c c  c c  c c }
\toprule
\multicolumn{1}{c}{\multirow{1}{*}{Method}} 
& \multicolumn{4}{c}{\multirow{1}{*}{BUTQDB Dataset}}  & \multicolumn{4}{c}{\multirow{1}{*}{{\edit{Private ECG}} Dataset}} \\ \cmidrule(lr){2-5}  \cmidrule(lr){6-9} 
\multicolumn{1}{c}{\multirow{1}{*}{ }}
& \multicolumn{2}{c}{\multirow{1}{*}{Level 2}}  & \multicolumn{2}{c}{\multirow{1}{*}{Level 3}} & \multicolumn{2}{c}{\multirow{1}{*}{Level 2}}  & \multicolumn{2}{c}{\multirow{1}{*}{Level 3}} \\ 
\cmidrule(lr){2-3} \cmidrule(lr){4-5}  \cmidrule(lr){6-7}   \cmidrule(lr){8-9}
\multicolumn{1}{c}{ } & \multicolumn{1}{c}{AUROC} & \multicolumn{1}{c}{AUPRC} & \multicolumn{1}{c}{AUROC} & \multicolumn{1}{c}{AUPRC} & \multicolumn{1}{c}{AUROC} & \multicolumn{1}{c}{AUPRC} & \multicolumn{1}{c}{AUROC} & \multicolumn{1}{c}{AUPRC} \\ \midrule
CAE + recons & 73.25{\scriptsize $\pm$0.20} & \textbf{75.46{\scriptsize $\pm$0.50}} & 84.43{\scriptsize $\pm$0.40} & 86.06{\scriptsize $\pm$0.30} & 67.78{\scriptsize $\pm$3.31} & 60.28{\scriptsize $\pm$2.54} & 79.96{\scriptsize $\pm$0.78} & 66.62{\scriptsize $\pm$0.86} \\
CVAE + log prob & 57.63{\scriptsize $\pm$1.50} & 59.26{\scriptsize $\pm$0.50} & 67.85{\scriptsize $\pm$1.50} & 81.10{\scriptsize $\pm$0.60} & 54.13{\scriptsize $\pm$0.74} & 52.91{\scriptsize $\pm$1.06} & \textbf{82.00{\scriptsize $\pm$0.09}} & 72.62{\scriptsize $\pm$0.07} \\ 
 \cmidrule(lr){1-9} 
Ours (m = 1) & 80.52{\scriptsize $\pm$1.35} & 76.96{\scriptsize $\pm$1.68} & 94.02{\scriptsize $\pm$0.71} & 91.84{\scriptsize $\pm$1.28} & 68.23{\scriptsize $\pm$1.40} & 62.84{\scriptsize $\pm$1.39} & 67.30{\scriptsize $\pm$0.81} & 57.61{\scriptsize $\pm$0.62}\\
Ours (m = 2) & 81.40{\scriptsize $\pm$1.45} & 77.28{\scriptsize $\pm$1.63} & 96.13{\scriptsize $\pm$0.39} & 94.42{\scriptsize $\pm$0.89} & 71.19{\scriptsize $\pm$3.75} & 66.44{\scriptsize $\pm$5.62} & 73.91{\scriptsize $\pm$4.96} & 66.45{\scriptsize $\pm$7.02}\\
Ours (m = 3) & 77.88{\scriptsize $\pm$0.96} & 73.55{\scriptsize $\pm$1.92} & 96.57{\scriptsize $\pm$0.53} & 95.05{\scriptsize $\pm$1.17} & 72.47{\scriptsize $\pm$2.71} & 67.72{\scriptsize $\pm$3.99} & 77.11{\scriptsize $\pm$5.51} & 68.61{\scriptsize $\pm$8.32}\\
Ours (m = 4) & 77.03{\scriptsize $\pm$1.38} & 73.31{\scriptsize $\pm$1.61} & 95.98{\scriptsize $\pm$0.67} & 94.60{\scriptsize $\pm$1.07} & 73.88{\scriptsize $\pm$2.80} & 71.23{\scriptsize $\pm$3.69} & 79.31{\scriptsize $\pm$5.86} & 74.29{\scriptsize $\pm$10.93}\\
Ours (m = 5) & 76.63{\scriptsize $\pm$1.16} & 72.93{\scriptsize $\pm$1.41} & 95.59{\scriptsize $\pm$0.94} & 94.20{\scriptsize $\pm$1.28} & 73.93{\scriptsize $\pm$2.71} & 70.49{\scriptsize $\pm$3.40} & 80.05{\scriptsize $\pm$4.69} & 76.61{\scriptsize $\pm$8.48}\\
Ours (m = 6) & 75.56{\scriptsize $\pm$1.33} & 70.92{\scriptsize $\pm$0.95} & 95.59{\scriptsize $\pm$1.21} & 94.03{\scriptsize $\pm$1.45} & 74.48{\scriptsize $\pm$2.88} & 72.12{\scriptsize $\pm$3.96} & 81.43{\scriptsize $\pm$5.09} & 77.48{\scriptsize $\pm$9.28}\\
Ours (m = 7) & 73.87{\scriptsize $\pm$2.72} & 69.04{\scriptsize $\pm$3.13} & 94.97{\scriptsize $\pm$2.26} & 93.10{\scriptsize $\pm$3.16} & 73.79{\scriptsize $\pm$2.61} & 71.28{\scriptsize $\pm$4.04} & 82.22{\scriptsize $\pm$4.66} & 78.71{\scriptsize $\pm$9.25}\\
Ours (m = 8) & 73.22{\scriptsize $\pm$0.85} & 68.38{\scriptsize $\pm$0.83} & 94.55{\scriptsize $\pm$0.64} & 92.78{\scriptsize $\pm$0.89} & 75.55{\scriptsize $\pm$2.55} & 74.11{\scriptsize $\pm$4.07} & 81.51{\scriptsize $\pm$5.02} & 77.88{\scriptsize $\pm$9.58}\\
Ours (m = 9) & 73.98{\scriptsize $\pm$1.06} & 69.36{\scriptsize $\pm$1.63} & 95.33{\scriptsize $\pm$1.09} & 93.78{\scriptsize $\pm$1.58} & 74.75{\scriptsize $\pm$3.48} & 73.47{\scriptsize $\pm$5.39} & 81.89{\scriptsize $\pm$5.33} & 78.98{\scriptsize $\pm$10.17}\\
Ours (m = 10) & 73.31{\scriptsize $\pm$1.96} & 68.22{\scriptsize $\pm$2.55} & 94.83{\scriptsize $\pm$2.42} & 92.85{\scriptsize $\pm$3.71} & 74.82{\scriptsize $\pm$3.19} & 72.96{\scriptsize $\pm$4.92} & 81.46{\scriptsize $\pm$5.26} & 78.26{\scriptsize $\pm$9.85}\\ [0.5ex] \cdashline{1-9}[1pt/1pt] \noalign{\vskip 0.5ex}
Ours (Ensemble) & \textbf{77.82{\scriptsize $\pm$1.05}} & 73.49{\scriptsize $\pm$1.31} & \textbf{95.84{\scriptsize $\pm$0.70}} & \textbf{94.25{\scriptsize $\pm$1.13}} & \textbf{74.28{\scriptsize $\pm$2.18}} & \textbf{72.09{\scriptsize $\pm$3.09}} & 80.25{\scriptsize $\pm$4.92} & \textbf{76.60{\scriptsize $\pm$9.84}}\\
 \bottomrule
\end{tabular}}
\vspace{1mm}
\caption{Noisy ECG signals detection performance of our approach on BUTQDB and Private ECG datasets measured by AUROC and AUPRC. The models are trained on Level $1$ samples and evaluated for Noisy ECG signals detection on Level $2$ and $3$ of the respective datasets. The results are averaged across $5$ seeds. We compare the Noisy ECG signals detection performance of Ours (Ensemble) with the baselines (CAE + recons, VAE + log prob) and observe that our approach obtains superior results (indicated by \textbf{bold} numbers).}
\vspace{-1mm}
\label{tab:train_from_scratch}
\end{table*}

\section{Methodology}
\label{methodology}
We present a two-stage framework for the detection of noisy ECG signals.

\vspace{1.5mm}
\parahead{Stage 1} We train a one-dimensional convolutional auto-encoder (CAE) model $g$ on a training dataset $P_\mathbf{X}$ composed of clean ECG samples.
The encoder is composed of one-dimensional convolutional neural networks (1D CNNs), and it extracts the latent features smaller than the input sample $\mathbf{x}$. The decoder comprises of transposed 1D CNNs, and it reconstructs output similar to the input sample using the latent features extracted from the encoder. The CAE model is trained to minimize the reconstruction error between the input sample and the reconstructed output generated by the decoder. 

\vspace{1.5mm}
\parahead{Stage 2} 
Following ~\cite{sehwag2021ssd, Xiao2021DoWR, Dua2022TaskAA}, we employ a cluster-conditioned detection method in the feature space to detect the noisy samples. Given the training dataset $P_\mathbf{X}$ composed of clean ECG samples, we obtain their features using the feature extractor $f$ (parameterized by the encoder of the convolutional auto-encoder $g$) that extracts the latent features of an input sample $\mathbf{x}$ from the CAE model $g$ trained on $P_\mathbf{X}$. Then, we split the features of the training data $P_\mathbf{X}$ into $m$ clusters using Gaussian Mixture Model (GMM).
Next, we model features in each cluster independently as multivariate gaussian distribution and use the Mahalanobis distance to calculate the Noise detection score $s({\mathbf{x}_{test}})$ of a test sample $\mathbf{x}_{test}$ as follows:
\begin{equation}
    s({\mathbf{x}_{test}}) = -\min \limits_{m} (f(\mathbf{x}_{test}) - \mathcal{\mu}_{m})^{T} \Sigma_{m} ^{-1} (f(\mathbf{x}_{test}) - \mathcal{\mu}_{m}),
    \label{eq:mahala}
\end{equation}
where $f(\mathbf{x}_{test})$ denotes the features of a test sample $\mathbf{x}_{test}$. $\mathcal{\mu}_{m}$ and $\Sigma_{m}$ denote the sample mean and sample covariance of the features of the $m^{\text{th}}$ cluster. Note that we negate the sign to align with the conventional notion that $s({\mathbf{x}_{test}})$ is higher for samples from training distribution and
lower for samples from other distributions (noisy samples). \editr{Essentially, we use the Noise detection score $s({\mathbf{x}_{test}})$ to detect the samples located away from the training data (Level 1 samples) in the latent space as noisy.}

\vspace{1.5mm}
\parahead{Ensembling}
We also present an ensembling approach in which we obtain Noise detection score $s({\mathbf{x}_{test}})$ for different values of $m \in [m_{1}, m_{2}, m_{3}, ..., m_{n}]$ and average them to obtain a final  score, $s_{\text{ensemble}}({\mathbf{x}_{test}})$. More specifically, we run the GMM model multiple times with different values of the number of clusters to split the features of the training dataset into clusters. For each run, we use Mahalanobis distance to calculate the Noise detection score. Then, we obtain the aggregated score by averaging the noise detection scores obtained from different runs.
This approach ensures a sample is detected as Noisy only if the majority of the participants in the ensemble agree with one another.

\section{Experiments and Results}
\label{sec:results}

\subsection{Datasets}
We evaluate the performance of the presented approach on two datasets.

\vspace{1.5mm}
\parahead{BUTQDB Dataset} Brno University of Technology ECG Quality Database (BUTQDB)\cite{BUTQDB} is a publicly available dataset for evaluating the ECG quality provided by Physionet. The samples are measured by using a mobile ECG recorder from $15$  subjects in balanced gender and age. Each sample has a noise level annotated by ECG experts.

\vspace{1.5mm}
\parahead{Private ECG Dataset} This dataset is collected in Asia from $2927$  subjects. The data covers a wide range of ages, $18$ at the youngest and $102$  at the oldest. This data also has a balanced gender ratio.

Both datasets follow the same criteria in annotating the noise level as described below \cite{BUTQDB}:

\begin{enumerate}
    \item Level $1$ : All significant waveforms (P wave, T wave, and QRS complex) are clearly visible, and the onsets and offsets of these waveforms can be detected reliably.
    \item Level $2$ : The noise level is increased, and significant points in the ECG are unclear (for example, PR interval and/or QRS duration cannot be measured reliably), but QRS complexes are clearly visible, and the signal enables reliable QRS detection. 
    \item Level $3$ : QRS complexes cannot be detected reliably, and the signal is unsuitable for any analysis. 
\end{enumerate}
\cref{fig:ecg_samples} presents ECG samples from Level $1$ , $2$ , and $3$ from the BUTQDB and Private ECG datasets. Level $2$ and Level $3$ samples looks much different from Level $1$  samples. Further, Level $3$ samples are more shifted away from Level $1$  samples relative to Level $2$ samples.

Level $1$ in BUTQDB and Private ECG datasets are composed of $22,828$ and $1,314$ samples, respectively.
We randomly split the Level $1$ dataset to make a $80/10/10\%$ train/val/test split. Note that the data of the same samples could be present in multiple splits. We use the BUTQDB dataset and form Level $2$ and Level $3$ noisy datasets containing $4568$ and $1150$ ECG samples, respectively. Further, we use the Private ECG dataset and form Level $2$ and Level $3$ noisy datasets, each containing $172$ ECG samples. All these noisy datasets are balanced and comprises of equal number of noisy and clean samples.

\begin{figure*}[!t]
\centering
     \includegraphics[width=0.9\textwidth]{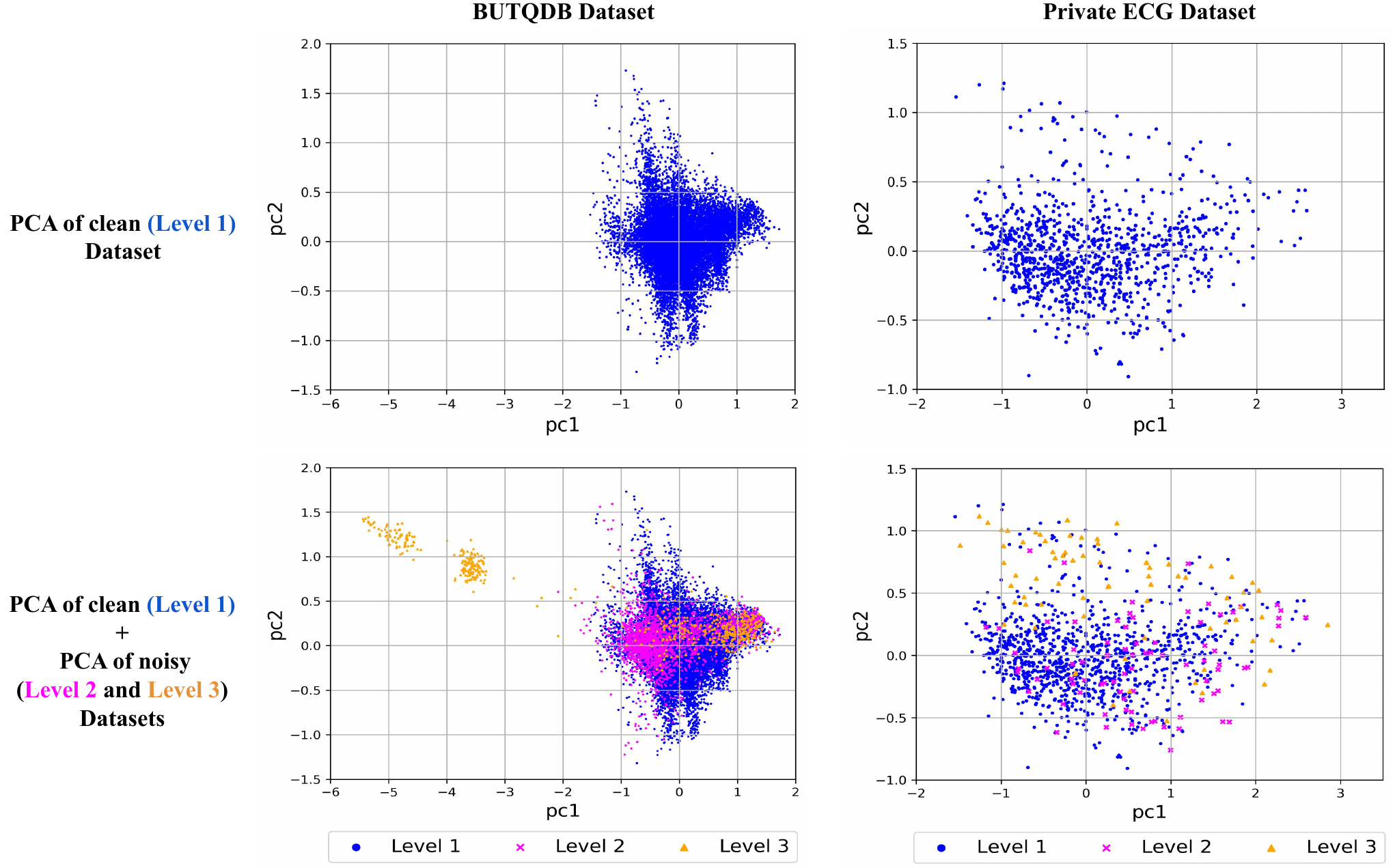}
    \vspace{-3mm}
    \caption{\emph{\textbf{(top)}} PCA visualization of clean (Level $1$) samples from BUTQDB and Private ECG datasets to demonstrate the location of clean samples. \emph{\textbf{(bottom)}} PCA visualization of clean (Level $1$) samples overlapped by PCA visualization of noisy samples (Level $2$ and Level $3$) from BUTQDB and Private ECG datasets to demonstrate the location of Level $2$ and Level $3$ samples relative to Level $1$  samples. In both the datasets,  Level
$3$ samples are more shifted from the Level $1$ samples compared
to Level $2$ samples}
     \label{fig:pca_visualization}
\end{figure*}

\begin{table*}[t]
\footnotesize
\centering
\resizebox{0.8\linewidth}{!}{
\begin{tabular}{c l c c c c c  c c c c c }
\toprule
\multicolumn{1}{c}{\multirow{1}{*}{ }}  & \multicolumn{1}{c}{\multirow{1}{*}{Method}} 
& \multicolumn{5}{c}{\multirow{1}{*}{Level 2}}  & \multicolumn{5}{c}{\multirow{1}{*}{Level 3}} \\ \cmidrule(lr){3-7}  \cmidrule(lr){8-12} 
& & \multicolumn{5}{c}{\multirow{1}{*}{Dataset Amount}}  & \multicolumn{5}{c}{\multirow{1}{*}{Dataset Amount}} \\ \cmidrule(lr){3-7}  \cmidrule(lr){8-12} 
& \multicolumn{1}{c}{ } & \multicolumn{1}{c}{20} & \multicolumn{1}{c}{40} & \multicolumn{1}{c}{60} & \multicolumn{1}{c}{80} & \multicolumn{1}{c}{100} & \multicolumn{1}{c}{20} & \multicolumn{1}{c}{40} & \multicolumn{1}{c}{60} & \multicolumn{1}{c}{80} & \multicolumn{1}{c}{100} \\ \midrule
\parbox[t]{2mm}{\multirow{9}{*}{\rotatebox[origin=c]{90}{AUROC}}}
& CAE + recons & 68.23 & 66.75 & 67.46 & 68.56 & 71.14 & \textbf{80.91} & 80.76 & 80.52 & 80.67 & 81.19\\
& CVAE + log prob & 53.66 & 47.70 & 52.03 & 51.66 & 54.99 & 78.42 & 77.92 & 79.57 &
79.22 & 81.88\\ \cmidrule(lr){2-12} 
& Ours (m = 1) & 71.34 & 71.24 & 71.01 & 70.98 & 69.12 & 69.94 & 68.85 & 68.85 & 68.65 & 67.70\\
& Ours (m = 2) & 70.89 & 72.26 & 73.70 & 73.40 & 73.00 & 68.66 & 73.32 & 75.08 & 76.20 & 76.28\\
& Ours (m = 3) & 70.82 & 74.96 & 76.28 & 74.82 & 72.01 & 68.48 & 80.67 & 78.87 & 77.27 & 79.34\\
& Ours (m = 4) & 70.80 & 77.11 & 77.39 & 77.11 & 75.38 & 68.66 & 86.78 & 81.96 & 81.72 & 82.95\\
& Ours (m = 5) & 74.81 & 78.22 & 77.58 & 76.95 & 75.53 & 77.53 & 88.72 & 86.11 & 81.52 & 85.24\\
& Ours (m = 6) & 74.85 & 78.37 & 79.26 & 79.41 & 75.70 & 78.02 & 89.02 & 86.83 & 84.46 & 84.90\\
& Ours (m = 7) & 74.50 & 76.81 & 77.58 & 79.75 & 77.81 & 78.41 & 85.88 & 84.55 & 85.32 & 88.30\\
& Ours (m = 8) & 75.28 & 78.26 & 81.14 & 79.91 & 75.69 & 78.98 & 91.25 & 90.82 & 89.41 & 88.03\\
& Ours (m = 9) & 69.74 & 78.26 & 81.58 & 79.42 & 76.46 & 71.34 & 91.25 & 89.82 & 86.47 & 87.11\\
& Ours (m = 10) & 71.67 & 78.26 & 81.15 & 78.46 & 78.42 & 70.66 & 91.25 & 91.08 & 84.63 & 90.27\\ [0.5ex] \cdashline{2-12}[1pt/1pt] \noalign{\vskip 0.5ex}
& Ours (Ensemble) & \textbf{72.05} & \textbf{78.91} & \textbf{79.85} & \textbf{78.5} & \textbf{76.31} & 72.47 & \textbf{89.43} & \textbf{88.05} & \textbf{84.11} & \textbf{86.24}\\

\midrule
\parbox[t]{2mm}{\multirow{9}{*}{\rotatebox[origin=c]{90}{AUPRC}}}
& CAE + recons & 59.69 & 58.72 & 59.44 & 60.49 & 62.93 & 67.79 & 67.75 & 67.45 & 67.61 & 68.07\\
& CVAE + log prob & 53.23 & 47.08 & 51.19 & 48.81 & 54.06 & 70.81 & 67.19 & 71.93 & 72.52 & 75.66\\  \cmidrule(lr){2-12}  
& Ours (m = 1) & 67.45 & 65.45 & 64.74 & 63.35 & 63.53 & 62.81 & 59.44 & 58.80 & 57.39 & 57.03\\
& Ours (m = 2) & 65.14 & 67.65 & 69.52 & 69.19 & 72.06 & 59.92 & 62.84 & 64.06 & 65.33 & 67.10\\
& Ours (m = 3) & 65.10 & 76.51 & 75.54 & 68.45 & 68.97 & 59.80 & 81.12 & 71.03 & 64.74 & 70.46\\
& Ours (m = 4) & 64.83 & 78.23 & 76.48 & 75.41 & 76.86 & 59.72 & 88.89 & 79.53 & 72.64 & 82.25\\
& Ours (m = 5) & 75.56 & 79.09 & 77.13 & 75.61 & 77.34 & 78.06 & 91.04 & 86.73 & 72.35 & 86.68\\
& Ours (m = 6) & 75.88 & 79.01 & 78.65 & 78.74 & 75.97 & 79.11 & 91.40 & 87.56 & 83.96 & 86.08\\
& Ours (m = 7) & 75.23 & 77.14 & 76.71 & 78.34 & 78.89 & 80.37 & 88.00 & 85.47 & 85.62 & 89.86\\
& Ours (m = 8) & 74.71 & 79.65 & 82.07 & 79.45 & 76.18 & 80.23 & 93.47 & 92.61 & 91.15 & 90.11\\
& Ours (m = 9) & 69.88 & 79.65 & 81.39 & 79.71 & 77.19 & 69.23 & 93.47 & 91.37 & 88.30 & 89.03\\
& Ours (m = 10) & 68.55 & 79.65 & 81.43 & 77.74 & 78.77 & 64.26 & 93.47 & 93.16 & 86.26 & 92.06\\ [0.5ex] \cdashline{2-12}[1pt/1pt] \noalign{\vskip 0.5ex}
& Ours (Ensemble) & \textbf{71.86} & \textbf{80.46} & \textbf{79.33} & \textbf{77.18} & \textbf{77.69} & \textbf{70.91 }& \textbf{91.84} & \textbf{89.74} & \textbf{84.24} & \textbf{88.07}\\

 \bottomrule
\end{tabular}}
\vspace{2mm}
\caption{Noisy ECG signals detection performance of our approach on Private ECG dataset measured by AUROC and AUPRC. The models pretrained on Level $1$ samples of BUTQDB dataset are finetuned on different amount of Level $1$ samples of private dataset and evaluated for Noisy ECG signals detection on Level $2$ and $3$ of the private dataset. We compare the Noisy ECG signals detection performance of Ours (Ensemble) with the baselines (CAE + recons, VAE + log prob) and observe that our approach obtains superior results (indicated by \textbf{bold} numbers).}
\vspace{-2mm}
\label{tab:finetune}
\end{table*}

\vspace{1.5mm}
\subsection{Experimental setup}
\vspace{1mm}
\parahead{Baselines} 
We compare our proposed method with two baselines.
\begin{itemize}
    \item \parahead{Convolutional Auto-encoder (CAE) + reconstruction error (AE + recons)} We determine the noisy samples based on the reconstruction loss produced by the trained Convolution Auto Encoder. The less the reconstruction loss, the more probability the given sample is a clean sample.
    \item \parahead{Convolutional Variational Auto-encoder + log probability (CVAE + log prob)} We determine the noisy samples based on the log probability calculated by the trained Convolutional Variational Auto-Encoder. 
The higher the log probability, the more likely the given sample is a clean sample.
\end{itemize}

\vspace{1.5mm}
\parahead{Evaluation Metrics}
We evaluate the effectiveness of our method for noisy ECG signals detection using two metrics, namely AUROC and AUPRC.
\begin{itemize}
    \item \parahead{AUROC} It measures the Area Under the Receiver Operating Characteristic curve, in which the true positive rate is plotted as a function of false positive rate for different threshold settings.
    \item \parahead{AUPRC-Out} It measures the Area Under the Precision-Recall (PR) curve. In a PR curve, the precision is plotted as a function of recall for different threshold settings. For the AUPRC-Out metric, the noisy samples are specified as positive.
\end{itemize}

\vspace{1.5mm}
\parahead{Training Details} 
We stacked two Convolutional layers for each encoder and decoder in CAE with latent hidden channel size of 64. All experiments are conducted with one $3090$ RTX GPU. We trained our model using AdamW optimizer with a learning rate of $1e-4$.  

\subsection{Results}
\vspace{1mm}
\parahead{Quantitative Results} 
\cref{tab:train_from_scratch} presents the noisy ECG signals detection performance of our presented method on BUTQDB and Private ECG datasets. 
We observe that our approach with different values of the number of clusters and ensembling approach outperforms the baselines for detecting Level $2$ and Level $3$ noisy samples across the two datasets. Although detecting the Level $2$ samples is challenging, the presented framework effectively detects them and outperforms the baselines. 
We also observe that all the methods, including baselines and our approach, obtain higher scores for Level $3$ samples than Level $2$ samples. This indicates that all methods detect samples from Level $3$ as noisier than samples from Level $2$, which aligns with the description of Level $2$ and Level $3$ samples.

\vspace{1.5mm}
\parahead{PCA Analysis} We analyze the PCA visualizations of the clean and noisy samples.
As shown in \cref{fig:pca_visualization}, we apply principal component analysis (PCA) on the feature representations obtained from the latent space of the convolutional auto-encoder (CAE) model to visualize the location of clean samples (Level $1$  samples) and noisy samples (Level $2$ and Level $3$).
We observe that Level $2$ samples are located close to the clean (Level $1$) samples whereas Level $3$ samples are located away from the clean (Level $1$) samples. Since the Level $3$ samples are more shifted from the Level $1$  samples compared to Level $2$ samples, our approach obtains higher AUROC and AUPRC scores for Level $3$ samples.

\vspace{1.5mm}
\parahead{Transfer to Private ECG Dataset} 
In the real-world, it is time-consuming and requires clinicians' effort to collect large datasets with annotations of Level $1$, $2$, and $3$ . 
To this end, we also studied whether the
presented approach, trained on the BUTQDB dataset, can be used to detect noisy samples in other datasets for which we have limited clean (Level $1$) samples. We use the convolutional auto-encoder (CAE) model trained on the BUTQDB dataset and finetune it on different percentages of Level $1$ samples from the Private ECG dataset. We finetune the model on $20\%, 40\%, 60\%, 80\%,$ and $ $1$ 00\%$ of the total Level $1$ samples in Private ECG dataset. Then, we evaluate the finetuned model for noisy ECG signals detection on Level $2$ and Level $3$ samples from the Private ECG dataset. 
\editr{Note that we do not test the generalization ability of our framework by training the model on the BUTQDB dataset and evaluating it on the Private ECG dataset (i.e., zero-shot test). This is because the Level $1$ samples of the two datasets might have different distributions due to diverse factors such as different dataset collection setup, ethnicity, etc. Due to this, Level $1$ samples of the Private ECG dataset might also be detected as noisy.}
\cref{tab:finetune} shows the noisy signals detection performance of the model finetuned on different amounts of Level $1$  samples from the Private ECG dataset. We observe that finetuning the model on a very small dataset hurts performance. In our experiments, finetuning on more than $40\%$ of dataset demonstrates decent noise detection performance of Level $2$ and Level $3$ noisy samples. Our finetuned model outperforms the baselines finetuned on BUTQDB.
Further, from \cref{tab:train_from_scratch} and \cref{tab:finetune}, we observe that finetuning the model pretrained on BUTQDB helps in better detection of Level $2$ and $3$ samples compared to training from scratch on Private ECG dataset.

\vspace{-1mm}
\section{Conclusion}
\label{sec:conclusion}
In this work, we attempt to bridge the gap between noisy ECG signals detection and deep learning and aimed to automatically detect the ECG signals contaminated by noise artifacts as they can affect the diagnosis. We present a two-stage deep learning-based method to detect noisy ECG samples. We conducted exhaustive experiments on two different ECG datasets composed of slightly noisy and highly noisy ECG samples to demonstrate the effectiveness of the two-stage framework in detecting the diverse noisy ECG samples. We also present a PCA analysis to justify the performance of the framework. To further bolster the effectiveness of the framework, we studied the transfer of the framework trained on one dataset to another dataset and observed that the framework can still identify the noisy samples. We hope that our work will open up a broader discussion around automatic detection of noisy ECG signals.

\section{Acknowledgments}
\label{sec:acknowledgement}
This work was supported by Institute of Information \& Communications Technology Planning \& Evaluation (IITP) grant (No.2019-0-00075, Artificial Intelligence Graduate School Program(KAIST)) funded by the Korea government (MSIT) and by Medical AI Inc.

\clearpage
\bibliographystyle{IEEEtran}
\bibliography{MAIN}






%
\end{document}